\documentstyle[aps,preprint,tighten,epsfig,prl]{revtex} % single-col/spacing
\newcommand{\ett}{\mbox{$E_T$}}
\newcommand{\dyrad}{\mbox{\tt{DYRAD}}}
\newcommand{\resbos}{\mbox{\tt{REBOS}}}
\newcommand{\ptt}{\mbox{$p_T$}}
\newcommand{\met}{\mbox{${\not\!\!E_T}$}}

\newcommand{\sgeq} {\raisebox{-.6ex}{${\textstyle\stackrel{>}{\sim}}$}}
\font\eightit=cmti8
\def\r#1{\ignorespaces $^{#1}$}
\begin{document}
\draft
%%\preprint{\begin{minipage}[t]{3in} 
%%		\flushright
%%		FERMILAB-PUB-98/256-E
%%		\end{minipage} }

%
\title{Measurement of the Lepton Charge Asymmetry in $W$-boson Decays
       Produced in $p\overline{p}$ Collisions}
\author{
\begin{sloppypar}
\noindent
F.~Abe,\r {17} H.~Akimoto,\r {39}
A.~Akopian,\r {31} M.~G.~Albrow,\r 7 A.~Amadon,\r 5 S.~R.~Amendolia,\r {27} 
D.~Amidei,\r {20} J.~Antos,\r {33} S.~Aota,\r {37}
G.~Apollinari,\r {31}
% ...
\end{sloppypar}
\vskip .026in
\begin{center}
(CDF Collaboration)
\end{center}
\vskip .026in
\begin{center}
\r 1  {\eightit Argonne National Laboratory, Argonne, Illinois 60439} \\
\r 2  {\eightit Istituto Nazionale di Fisica Nucleare, University of Bologna,
I-40127 Bologna, Italy} \\
\r 3  {\eightit Brandeis University, Waltham, Massachusetts 02254} \\
\r 4  {\eightit University of California at Los Angeles, Los 
Angeles, California  90024} \\  
\r 5  {\eightit University of Chicago, Chicago, Illinois 60637} \\
\r 6  {\eightit Duke University, Durham, North Carolina  27708} \\
\r 7  {\eightit Fermi National Accelerator Laboratory, Batavia, Illinois 
60510} \\
\r 8  {\eightit University of Florida, Gainesville, FL  32611} \\
\r 9  {\eightit Laboratori Nazionali di Frascati, Istituto Nazionale di Fisica
               Nucleare, I-00044 Frascati, Italy} \\
\r {10} {\eightit University of Geneva, CH-1211 Geneva 4, Switzerland} \\
\r {11} {\eightit Harvard University, Cambridge, Massachusetts 02138} \\
\r {12} {\eightit Hiroshima University, Higashi-Hiroshima 724, Japan} \\
\r {13} {\eightit University of Illinois, Urbana, Illinois 61801} \\
\r {14} {\eightit Institute of Particle Physics, McGill University, Montreal 
H3A 2T8, and University of Toronto,\\ Toronto M5S 1A7, Canada} \\
\r {15} {\eightit The Johns Hopkins University, Baltimore, Maryland 21218} \\
\r {16} {\eightit Institut f\"{u}r Experimentelle Kernphysik, 
Universit\"{a}t Karlsruhe, 76128 Karlsruhe, Germany} \\
\r {17} {\eightit National Laboratory for High Energy Physics (KEK), Tsukuba, 
Ibaraki 305, Japan} \\
\r {18} {\eightit Ernest Orlando Lawrence Berkeley National Laboratory, 
Berkeley, California 94720} \\
\r {19} {\eightit Massachusetts Institute of Technology, Cambridge,
Massachusetts  02139} \\   
\r {20} {\eightit University of Michigan, Ann Arbor, Michigan 48109} \\
\r {21} {\eightit Michigan State University, East Lansing, Michigan  48824} \\
\r {22} {\eightit University of New Mexico, Albuquerque, New Mexico 87131} \\
\r {23} {\eightit The Ohio State University, Columbus, OH 43210} \\
\r {24} {\eightit Osaka City University, Osaka 588, Japan} \\
\r {25} {\eightit Universita di Padova, Istituto Nazionale di Fisica 
          Nucleare, Sezione di Padova, I-35131 Padova, Italy} \\
\r {26} {\eightit University of Pennsylvania, Philadelphia, 
        Pennsylvania 19104} \\   
\r {27} {\eightit Istituto Nazionale di Fisica Nucleare, University and Scuola
               Normale Superiore of Pisa, I-56100 Pisa, Italy} \\
\r {28} {\eightit University of Pittsburgh, Pittsburgh, Pennsylvania 15260} \\
\r {29} {\eightit Purdue University, West Lafayette, Indiana 47907} \\
\r {30} {\eightit University of Rochester, Rochester, New York 14627} \\
\r {31} {\eightit Rockefeller University, New York, New York 10021} \\
\r {32} {\eightit Rutgers University, Piscataway, New Jersey 08855} \\
\r {33} {\eightit Academia Sinica, Taipei, Taiwan 11530, Republic of China} \\
\r {34} {\eightit Texas A\&M University, College Station, Texas 77843} \\
\r {35} {\eightit Texas Tech University, Lubbock, Texas 79409} \\
\r {36} {\eightit Istituto Nazionale di Fisica Nucleare, University of Trieste/
Udine, Italy} \\
\r {37} {\eightit University of Tsukuba, Tsukuba, Ibaraki 315, Japan} \\
\r {38} {\eightit Tufts University, Medford, Massachusetts 02155} \\
\r {39} {\eightit Waseda University, Tokyo 169, Japan} \\
\r {40} {\eightit University of Wisconsin, Madison, Wisconsin 53706} \\
\r {41} {\eightit Yale University, New Haven, Connecticut 06520} \\
\end{center}
\date{\today}
}

\maketitle
\begin{abstract}
We describe a measurement of the charge asymmetry of leptons from $W$
boson decays in the rapidity range $0<|y_l|<2.5$ using
$W{\rightarrow}~e\nu,~\mu\nu$ events from $110\pm7 $~pb$^{-1}$ of data
collected by the CDF detector during 1992-95.
The asymmetry data constrain the ratio of $d$ and $u$ quark momentum
distributions in the proton over the $x$ range of $0.006$ to $0.34$ at
$Q^2 \approx M_W^2$.  The asymmetry predictions that use parton
distribution functions obtained from previously published
CDF data in the central rapidity region ($0.0<|y_l|<1.1$) do not agree with
the new data in the large rapidity region ($|y_l|>1.1$).
\end{abstract}
\vspace{0.2in}
\vspace{0.2in}

\narrowtext

At Tevatron energies, $W^+$ ($W^-$) bosons are produced in
$p\overline{p}$ collisions primarily by the annihilation of $u$ ($d$)
quarks in the proton and $\overline{d}$ ($\overline{u}$) quarks from the
antiproton. Because $u$ quarks carry on average more momentum than $d$
quarks~\cite{MRSA,CTEQ3M,GRV94}, ~the $W^+$ bosons tend to follow the
direction of the incoming proton and the $W^-$ bosons that of the antiproton.
The charge asymmetry in the production of $W$ bosons is related to the $u$
and $d$ quark distributions at $Q^2 \approx M_W^2$, and is roughly
proportional~\cite{ELB,ADM} to the ratio of the difference to the sum
of the quantities $\frac{d}{u}(x_1)$ and $\frac{d}{u}(x_2)$, 
where $x_1$ and $x_2$ are
the fractions of the nucleon's momentum carried by the quarks in the $p$ and
$\overline{p}$, respectively.  $\frac{d}{u}(x)$ is the ratio of the $d$ to $u$
quark parton distribution functions for quarks carrying a momentum
fraction, $x$, of the nucleon's momentum.

$W$ asymmetry measurements~\cite{Old_ASYM} from 20~pb$^{-1}$ of data
collected by the Collider Detector at Fermilab (CDF) in 1992-93
demonstrated the measurement to be more sensitive than deep inelastic
scattering experiments to the $\frac{d}{u}$ ratio within the $x$ range
accessible by the $W$ data at the Tevatron.
These data contribute to global analyses used to extract parton
distribution functions (PDFs) in the nucleon~\cite{MRSA,CTEQ3M,GRV94}. 
These functions are used in the calculation
of all hadronic cross sections.  CDF's previous asymmetry
data~\cite{Old_ASYM} improved the understanding of the $u$ and $d$ quark
momentum distributions in the proton in the $x$ range
$0.007<x<0.24$. The use of these data
resulted in a reduction in the uncertainty from PDFs in the $W$ mass
measurement~\cite{Wmass} from $\approx$ 75 to $\approx$ 50~MeV/c$^2$.  A
further reduction of this error to $\approx$\nobreak~20~MeV/c$^2$ is
possible~\cite{Wmass2} using the new data presented here. Recently fixed
target Drell-Yan data~\cite{FT_DY} has also been used to constrain the
($u-d$) sea quark distribution.

In this letter, we describe the asymmetry results at
\mbox{$\sqrt{s}=1.8$~TeV} using data from the entire 1992-95 period,
using a total integrated luminosity of $110\pm7 $~pb$^{-1}$ -- a 5-fold
increase relative to the 1992-93 data. The new measurements agree with
and supersede the previous results. The asymmetry measurement is
extended to larger rapidity~\cite{met_def} ($|y_l|<2.4$) by the 
introduction of a new
charge determination technique~\cite{RATIO} for electrons, and also by
using data from the forward muon detector ($1.9<|y_l|<2.5$).  These
electron and muon data at large rapidity provide information about PDFs
for a larger $x$ range ($0.006<x<0.34$) than previously available. The
mean $x$ values of the $d$ and $u$ parton distributions probed
by the CDF asymmetry data are shown in Table ~\ref{xbins}.

Since the $W$ rapidity is experimentally undetermined because of the
unknown longitudinal momentum of the neutrino from the $W$ decay, we
measure the lepton charge asymmetry which is a convolution of the $W$
production charge asymmetry and the $V-A$ asymmetry from the 
$W$ decay.  The two asymmetries tend to cancel at large values 
($|y_l|~\sgeq~2$) of
rapidity. 
However, if one assumes $W$ decays proceed via a pure $V-A$
interaction, the lepton asymmetry is still sensitive to the parton
distribution functions.
The lepton charge asymmetry is defined as:
\begin{equation}
A(y_l)=\frac{d\sigma^+/dy_l-d\sigma^-/dy_l}
            {d\sigma^+/dy_l+d\sigma^-/dy_l},
\end{equation}
where $d\sigma^+$ ($d\sigma^-$) is the cross section for $W^+$~($W^-$)
decay leptons as a function of lepton rapidity, with positive rapidity
being defined in the proton beam direction.  If the detection
efficiencies and acceptances for $l^+$ and $l^-$ are equal, then the
uncertainties on these quantities do not affect $A(y_l)$.  

The tracking detectors at CDF~\cite{CDF} are the Vertex Time
Projection Chambers (VTX), the Central Tracking Chamber (CTC,
$|\eta|<1.8$), the Silicon Vertex Detector (SVX, $|\eta|<2.3$), the
Central Muon Chambers (CMU and CMX, $|\eta|<1.0$), and the Forward Muon
Detector (FMU, $1.9<|\eta|<3.5$).  The VTX provides $r$-$z$ tracking
information out to a radius of 22 cm for $|\eta|<3.25$ and is used to
measure the $z$ location~\cite{met_def} of the primary interaction
vertex. The CTC is an 84 layer drift chamber inside the 1.4 T solenoidal
magnet which provides the curvature measurement of electrons and
muons from $W$ decays.  The CTC has a momentum resolution of
$\delta(1/{p_T}) = 0.00081$~(GeV/c)$^{-1}$~\cite{Wmass}.
The SVX is a 51 cm long, 4 layer precision
vertex detector with a track position resolution~\cite{RATIO} of
$\approx 10$ $\mu m$. Because $p\overline{p}$ interactions at CDF are
spread along the beamline with an RMS of $\approx30$~cm, the
geometrical acceptance of the SVX is about 60\% for $W$ candidate events.
In this analysis, the SVX complements the CTC for electron tracking at
high $|\eta|$ ($1.2<|\eta|<2.3$). The FMU consists of two steel magnets
(toroids) alternately inserted between 3 drift chambers.

$W$ decays to leptons are identified by a charged track pointing either at hits
in the muon chambers or a cluster of energy in the electromagnetic
calorimeters (EM) accompanied by large missing transverse energy
({\met})~\cite{met_def}. Electron candidates are required to
fall within the fiducial regions of either the central ($|y_l|<1.1$) or
the plug ($1.1<|y_l|<2.4$) EM calorimeters and to pass identification
cuts based on the EM shower's profile determined with test beam
electrons. Muon candidates are required to have a track in the muon
tracking system and a minimum ionizing particle signal in the hadronic
and EM calorimeters.  The transverse energy ({\ett}) of the lepton and
\met\ are required to be greater than 25 GeV. To further
reduce the background from dijet events where one jet is incorrectly
reconstructed as a charged lepton and where a large \met\
results from shower fluctuations or uninstrumented regions in the
detector, events with a jet~\cite{jet} with $E_{T}^{\rm jet} > 30$ GeV
are rejected.

 The data are divided into five samples: central electrons
($|y_l|<1.1$), central muons ($|y_l|<1.0$), plug electrons within the
SVX fiducial region (plug-SVX, $1.1<|y_l|<2.4$), plug electrons outside
the SVX geometrical acceptance region but with a CTC track (plug-CTC,
$1.1<|y_l|<1.8$), and forward muons ($1.9<|y_l|<2.5$). In total 88047
events are selected.
A reliable charge determination and a good energy
measurement are essential for this analysis.  Central electrons, central
muons, and plug electrons outside the SVX are required to have an
associated CTC track. A reliable charge determination is
ensured by a CTC track curvature ($C$) significance requirement of
$C/\delta C>2$ (where $\delta C$ is the error in the curvature
measurement). The charge of forward muons is measured well in the FMU
toroidal magnets. The energy of plug electrons is
measured by the EM calorimeters, and the $\eta$ and $\phi$ are measured
by the tower segmentation~\cite{met_def}.  However, tracking information
is needed in order to determine the charge. The geometrical acceptance
of the plug calorimeters is only partially covered by the CTC. Therefore, the
efficiency for finding a CTC track, and hence determining the charge,
drops quickly as a function of rapidity and becomes zero for $|y_l|>1.8$.

A new charge determination technique~\cite{RATIO} is introduced to
identify the charge of plug electrons within the SVX geometrical
fiducial region.  At CDF, positively (negatively) charged particle 
trajectories are
bent in increasing (decreasing) $\phi$ inside the solenoidal magnetic
field. The SVX track stub measures precisely the initial track direction
in $\phi$. The plug EM calorimeter (PEM)~\cite{PEMNIM} shower centroid
measurement combined with the location of the vertex position yields
another measurement of $\phi$, and the sign of the difference of the two
$\phi$ measurements determines the charge.
The shower centroid is measured with a strip position detector placed at
shower maximum in the pseudo-rapidity region between $1.2-1.8$.  In the
pseudo-rapidity region between $1.8-2.4$, where there is no strip
detector coverage, the centroid is measured from the calorimeter tower
information in $\eta$ and $\phi$.
We define the ratio $R =
\delta\phi_{\rm {measured}}/\left|\delta\phi_{{\rm expected}}\right |$, where
$\delta\phi_{{\rm measured}}=\phi_{{\rm PEM}}-\phi_{{\rm SVX}}$, and
$\delta\phi_{{\rm expected}}$ is calculated from the calorimeter energy, the
radial location of the electron shower and the magnetic field strength.
$R$ peaks at +1 and --1 for positrons and electrons, respectively. The
measurement error on $R$ is 0.3, 0.5 and 0.8 for $|y_l|$ bins centered at
1.3, 1.8 and 2.2, respectively. This technique effectively doubles the
number of plug electrons that can be used in the measurement and extends
the charge measurement to $|y_l|=2.4$.

A possible charge or \ett\ dependence of each trigger is investigated
by using data from several independent triggers.  No evidence of charge
dependence is found in the electron and central muon data samples. The
forward muon data taken with the toroid polarities set to focus $\mu^+$
and $\mu^-$, respectively, are averaged to cancel out effects of any
charge bias in the trigger from geometrical acceptance and alignment
errors.  The efficiency of the plug electron triggers does not depend on
charge, but is not $100\%$ at \ett\ of 25 GeV. At \ett\ = 25~GeV the
average trigger efficiency for the plug-SVX sample is 78\%.
A correction is applied to the measured asymmetry in each rapidity
bin, using a Monte Carlo calculation and the measured trigger efficiency
as a function of \ett\ and $y_l$.  The correction to $A(y_l)$ is found
to be less than 0.005 compared with a typical statistical error of 0.015
in the PEM region.

Sources of charge bias in the event selection are investigated by
selecting high \ett\ leptons from either a sample of $Z$ or $W$ events,
which satisfy tight kinematic constraints. No charge-dependent
bias in acceptance or efficiencies has been found. The charge
misidentification rate for muons is negligible.  The electron charge
misidentification rate (e.g. due to multiple tracks from conversion of
Bremsstrahlung photons) in the central region is determined from the
rate of same sign central-central $Z$ electrons. In the plug region, the
rates are determined from $W$ electron data using the number of events
outside the two peaks in the $R$ distribution. This is checked against
the same sign
central-plug $Z$ electrons as a function of rapidity. The electron
charge misidentification rate ranges from $0.2\%$ in the central data
sample to $10\%$ in the largest rapidity bin ($|y_l|=2.20$) of the plug
electron data. The charge misidentification rate in the FMU sample is
$<$~1\%.
The effect of the charge misidentification acts to dilute the
charge asymmetry and is corrected on a bin by bin basis.

~~All backgrounds have been investigated and found to be small (see
Table~\ref{BKG}). The average $W \rightarrow \tau\nu$ background is 
2.0~$\pm$~0.2\% where the error accounts for the small rapidity dependence.
In the plug electron sample, the background from misidentified dijet
events is the largest. This background is charge-symmetric and dilutes
the asymmetry. In the central electron sample, the background from
$W \rightarrow \tau\nu \rightarrow e \nu\nu\nu$ is the largest.
For the central muon sample the largest background is misidentified
$Z\rightarrow\mu^+\mu^-$ where one of the muons is outside the
acceptance of the CTC. In the latter two cases, the backgrounds are not
charge-symmetric and the corrections are made using the asymmetry
estimated by Monte Carlo calculations. The background from misidentified
$Z$ decays to electrons is negligible because the EM calorimeters have a
much larger geometric acceptance than the muon chambers or the CTC. In
the forward muon sample the background from $Z$ events is 6.5\% and the
charge-symmetric background totals 8\% of which half is from dijets and
half from lower \ptt\ muons which are misreconstructed because of
random extra hit background in the forward drift chambers.
%% activity in the chambers unrelated to the primary interaction.
The $Z\rightarrow\tau^+\tau^-$ contamination is negligible. The
background from cosmic rays in the muon sample is $<$ 0.2\%. The
$A(y_l)$ values are corrected for the backgrounds on a bin by bin basis.

%%Figure~\ref{RAWA} shows the uncorrected asymmetry measurements.
%%Consistency
%%of the data from different detectors is expected because most
%%backgrounds and systematic errors are small. 

\begin{figure}
\epsfxsize=18cm
\epsfysize=15cm
\centerline{\epsffile{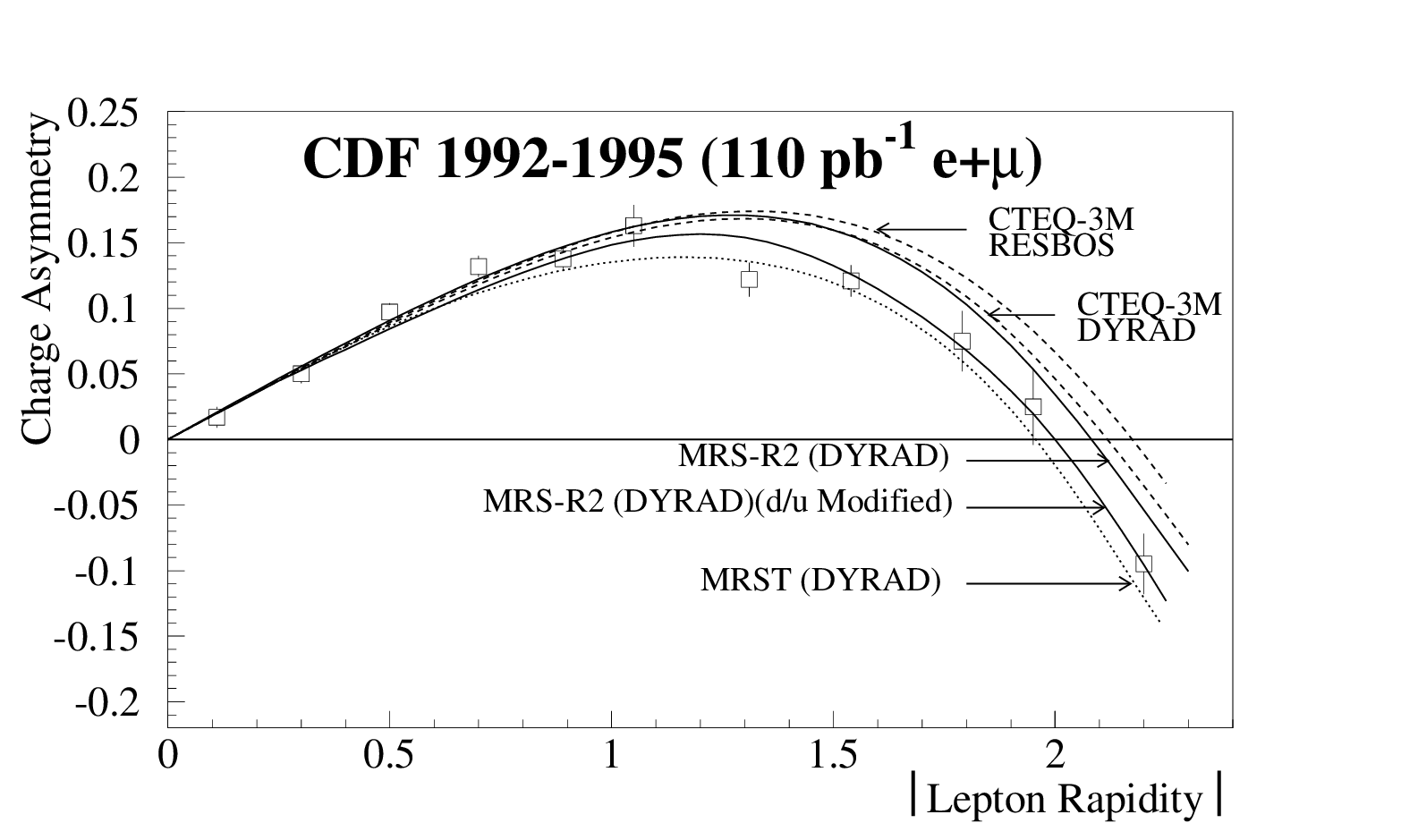}}
\caption{
 The fully corrected charge asymmetry. Data from all the detectors for
 positive and negative $y_l$ are combined. The statistical and
 systematic errors are added in quadrature.}
\label{ASYM} 
\end{figure}

By CP invariance the
asymmetry at positive $y_l$ is equal in magnitude and opposite in sign
to that at negative $y_l$ allowing the two measurements to be combined.
Figure~\ref{ASYM} shows
the fully corrected asymmetry after taking the weighted mean of the
various data sets and combining the positive and negative
$y_l$ bins.  The asymmetry and uncertainties are listed in Table~\ref{DATA}.
Figure~\ref{ASYM} shows the predictions of Quantum Chromodynamics (QCD)
calculated to Next-to-Leading-Order using the program
{\dyrad}~\cite{Dyrad} with several parameterizations of parton distribution
functions as input. The MRS-R2~\cite{MRSR} and
CTEQ-3M~\cite{CTEQ3M} PDFs have been extracted with the inclusion of the
1992-93 asymmetry data in their global fits (the 1992-93 data
constitutes $20\%$ of the sample reported here).  As shown in
Figure~\ref{ASYM}, the predictions using these PDFs are in good
agreement with the present data in
the central region ($|y_l|<1.1$), with $\chi^2$ per degree of freedom (d.o.f)
values in the range 0.7--1.3.
However, at high rapidity, these
predictions are generally higher than the data, with $\chi^2$/d.o.f values in
the range 5--10, indicating that the PDF
parameterizations should be modified in the range $0.006<x<0.34$ 
(see Table~\ref{xbins}). A recent re-analysis~\cite{NMC_REA} of NMC muon
scattering data~\cite{NMC_DATA} on hydrogen and deuterium with improved
corrections for nuclear binding effects in the deuteron has shown the
need for a correction to the $\frac{d}{u}$ ratio of the form : $\delta(\frac{d}{u}) =
0.1(x+x^2)$.  The prediction using the corrected MRS-R2 PDF is shown
in Figure~\ref{ASYM} and shows reasonable agreement with the measured
asymmetry at high $|y_l|$.  Recent global parton distribution
fits~\cite{NEWPDF} (e.g. MRST) that have incorporated the data
presented in this paper also lead to agreement at high $|y_l|$.
The theoretical uncertainties arising from the effect of the
finite charm quark mass~\cite{HEAVY} are much smaller than the
measurement errors.

The $W$ decay lepton charge asymmetry is mostly sensitive
to the $\frac{d}{u}$ ratio. However, at large rapidity it is also affected by the $W$
production \ptt\ spectrum.  The \dyrad\ calculation does not reproduce
the production \ptt\ spectrum of $W$ events at low {\ptt}.  Therefore, it
is necessary to check how well the \dyrad\ calculation compares to one
more suitable for the low $W$ \ptt\ region. 
The \dyrad\ prediction is compared to a
calculation including soft gluon resummation at all orders in
perturbation theory implemented in the program
{\resbos}~\cite{RESBOS}. Both programs yield identical results for the $W$
rapidity distribution, but different results for the $W$ \ptt\
distribution.  At Tevatron energies soft gluon radiation is mainly
responsible for the \ptt\ of $W$ and $Z$ bosons in the range of \ptt\ $ < 30$
GeV/c, and \resbos\ can reproduce the \ptt\ spectrum of $Z$ bosons~\cite{Z_PT}.
Figure~\ref{ASYM} shows the comparison between the asymmetry predictions of the
\dyrad\ and \resbos\ programs using CTEQ-3M~\cite{CTEQ3M} PDFs. 
The two calculations agree very well in the central region
($|y_l|<1.1$), which is the region used in the $W$ mass
determination~\cite{Wmass}.  The difference between the two calculations is
mainly at $|y_l|>1.7$, as illustrated in Figure~\ref{ASYM}.
 
In summary, the asymmetry data demonstrate the value of
collider data in the measurement of parton distribution functions and place
the strongest constraint on the $\frac{d}{u}$ ratio of quark momentum
distributions in the proton over the range $x$ of $0.006$ to $0.34$ at $Q^2
\approx M_W^2$. The data indicate the need for modifications in the 
$\frac{d}{u}$ ratio in the PDFs constrained by the previous CDF
asymmetry data.

We thank the Fermilab staff and the technical staffs of the
participating institutions for their vital contributions.  This work was
supported by the U.S. Department of Energy and National Science Foundation;
the Italian Istituto Nazionale di Fisica Nucleare; the Ministry of Education,
Science and Culture of Japan; the Natural Sciences and Engineering Research
Council of Canada; the National Science Council of the Republic of China; 
the A. P. Sloan Foundation; and the Swiss National Science Foundation.

\begin{table}
\caption{The mean $x$ values of the $d$ and $u$ parton distributions
 probed by the CDF asymmetry data for the $W^+$ boson as a function of
 the lepton rapidity. The $\frac{d}{u}$ ratios from the CTEQ-3M 
 parameterization are shown for these $x$ values. 
 The mean $x$ values for the $W^-$ at positive $y_l$ are the same as 
 those for the $W^+$ at the same negative $y_l$.}
\begin{center}
\begin{tabular}{lllll}  
  $\langle{y_l}\rangle$  & $\langle{x_u}\rangle$ & 
  $\langle{x_d}\rangle$  & $\frac{d}{u}(x_u)$ & 
  $\frac{d}{u}(x_d)$ \\[0.1cm] \hline
 $-$2.20  & 0.009& 0.219 &  0.911 & 0.412 \\ 
 $-$1.54  & 0.016& 0.129 &  0.864 & 0.542 \\ 
 $-$1.10  & 0.021& 0.088 &  0.837 & 0.625 \\ 
 $-$0.49  & 0.038& 0.051 &  0.765 & 0.722 \\ 
  0.00  & 0.044& 0.044 &  0.744 & 0.744 \\ 
  0.49  & 0.099& 0.021 &  0.601 & 0.837 \\ 
  1.10  & 0.165& 0.012 &  0.483 & 0.889 \\ 
  1.54  & 0.225& 0.009 &  0.405 & 0.911 \\ 
  2.20  & 0.335& 0.006 &  0.306 & 0.936 \\ 
\end{tabular}
\end{center}
\label{xbins}
\end{table}
\begin{table}
\caption{Backgrounds (\%) in the $W\rightarrow e\nu$ and
$W\rightarrow \mu \nu$ charge asymmetry event samples.}
\begin{center}
\begin{tabular}{ccc}  
Sample  & QCD dijets &
 $Z\rightarrow e^+e^-$ or $\mu^+\mu^-$ \\ \hline
Central $e$    & ${0.7\pm0.2}$ & $< 0.2$   \\
Central $\mu$  & ${0.6\pm0.2}$ & ${4.7\pm0.7}$  \\ 
Plug-SVX $e$   & ${1.6\pm0.3}$ & $< 0.2$    \\
Plug-CTC $e$   & ${2.4\pm0.6}$ & $< 0.2$    \\
Forward $\mu$  & ${4\pm2}$     & ${6.5\pm0.6}$ \\ 
\end{tabular}
\end{center}
\label{BKG}
\end{table}
\begin{table}
\caption{The charge asymmetries (after all corrections) 
and statistical and systematic uncertainties
in the combined $e$ and $\mu$ channels.}
\begin{center}
\begin{tabular}{crcc} 
 $\langle|y_l|\rangle$ & $A(y_l)$ & $\sigma_{{\rm sys}}$ 
& $\sigma_{{\rm stat+sys}}$\\ 
\hline
0.11 & 0.017 & $\pm0.000$ & $\pm0.008$\\
0.30 & 0.050 & $\pm0.000$ & $\pm0.007$\\
0.50 & 0.097 & $\pm0.000$ & $\pm0.007$\\
0.70 & 0.132 & $\pm0.001$ & $\pm0.008$\\
0.89 & 0.138 & $\pm0.001$ & $\pm0.009$\\
1.05 & 0.163 & $\pm0.002$ & $\pm0.016$\\
1.31 & 0.122 & $\pm0.002$ & $\pm0.013$\\
1.54 & 0.121 & $\pm0.003$ & $\pm0.012$\\
1.79 & 0.075 & $\pm0.005$ & $\pm0.023$\\
1.95 & 0.025 & $\pm0.010$ & $\pm0.029$\\
2.20 &$-$0.095 & $\pm0.015$ & $\pm0.023$\\
\end{tabular}
\end{center}
\label{DATA}
\end{table}
%
%%\begin{figure}
%%\epsfxsize=18cm
%%\epsfysize=15cm
%%\centerline{\epsffile{fig1.ps}}
%%\caption{
%%    The lepton charge asymmetry (prior to any corrections) as measured
%%    in each detector  (Central EM-CTC, Central Muon-CTC, Plug EM-SVX,
%%    Plug EM-CTC, and Forward Muon). The asymmetry predictions are
%%    calculated using the \dyrad\ program using CTEQ-3M
%%    parton distribution functions.}
%%\label{RAWA} 
%%\end{figure}
%

\begin{thebibliography}{999}
\bibitem{met_def}
In the CDF coordinate system, $\theta$ is the polar angle with respect
to the proton beam direction and $\phi$ is the azimuthal angle. The
pseudorapidity $\eta$ is defined as $-$ln~tan($\theta/2$) and the 
lepton rapidity ($y_l$) as ~ln$\frac{E+E_z}{E-E_z}$. 
Transverse energy is defined as  $E_T$ = $E~sin~\theta$, where $E$ is
the lepton energy.
The missing transverse energy, {$\met$}, is defined as the 
magnitude of the vector sum  of $E_T$ in all calorimeter cells with 
$|\eta|<3.6$, and including any muon {\ptt}.
\bibitem{jet}
The jet energy is clustered using a cone algorithm with radius 
${\Delta}R = \sqrt{\Delta\eta^2 + \Delta\phi^2} = 0.7$.
\end{thebibliography}
\end{document}